# Coarsened Lattice Spatial Disorder in the Thermodynamic Limit

R. Piasecki [1, a], A. Czaiński [2]

[1] *Institute of Chemistry, University of Opole, Oleska 48, PL 45052 Opole, Poland*
[2] *Institute of Physics, University of Opole, Oleska 48, PL 45052 Opole, Poland*

Much research effort has been devoted for developing simple tools to correlate the macroscopic properties of inhomogeneous media with their microstructure properties [1]. To characterize at every length scale a digitized image microstructure an adaptation of Shannon information entropy was used [2]. Recently, the normalized information entropy, where the initial entropy for the perfectly random system is subtracted from the actual entropy calculated for the real particle configuration has been considered [3]. The different idea of using a linear transformation of configurational entropy as a measure of the degree of inhomogeneity of the spatial distribution, mentioned in [4], was applied to point [5] and finite-sized [6] objects.

To investigate how it reveals an information concerning the changes of spatial disorder in the thermodynamic limit first we briefly recall the two formulas of [6]. When a binary image of size $L \times L$ in pixels is treated as a set of $n$ indistinguishable finite-sized objects, the black pixels represents "particles" distributed in $\chi = (L/k)^2$ lattice cells of size $k \times k$. The numbers $n_i$ of particles in $i$th cell fulfil the two constraints $n_1 + n_2 + ... + n_\chi = n$ and $n_i \leq k^2$. Using the notation of [6] the measure $S_\Delta(k)$ is obtained for every length scale $k$ by subtracting the actual configurational entropy $S(k)$ from the entropy $S_{\max}(k)$ related to the most spatially ordered particle configuration and averaging this difference over the number of cells $\chi$

$$S_\Delta(k) = \frac{r_0}{\chi} \ln\left[\frac{k^2 - n_0}{n_0 + 1}\right] + \frac{1}{\chi} \sum_{i=1}^{\chi} \ln\left[\frac{n_i!(k^2 - n_i)!}{n_0!(k^2 - n_0)!}\right], \qquad (1)$$

where $r_0 = n \bmod \chi$, $r_0 \in 0, 1, ..., \chi - 1$ and $n_0 = (n - r_0)/\chi$, $n_0 \in 0, 1, ..., k^2 - 1$ and the Boltzmann constant is set to $k_B = 1$. For the periodic pattern $mL \times mL$ the corresponding formula in the thermodynamic limit $m \to \infty$, i.e. for $k$ = const, $n' \equiv m^2 n \to \infty$ and $\chi' \equiv m^2 \chi \to \infty$ in such a way that $n'/\chi' k^2 \equiv \varphi$ = const, can be written as follows

$$S_\Delta(\varphi) \cong -k^2 \left\{ \varphi \ln \varphi + (1-\varphi) \ln(1-\varphi) - \sum_{\{n_i\}} F_{n_i}(\varphi) \left[ \varphi_{n_i} \ln \varphi_{n_i} + (1-\varphi_{n_i}) \ln(1-\varphi_{n_i}) \right] \right\}, \quad (2)$$

where $\varphi_{n_i} \equiv n_i / k^2$ refers to the local concentration, $F_{n_i}(\varphi) = \lim_{m \to \infty} [m^2 g_{n_i}(\varphi) / m^2 \chi] \equiv g_{n_i}(\varphi)/\chi$ denotes the frequency of occurrence of the cells with $n_i$ particles and $g_{n_i}(\varphi)$ is the number of such cells for the initial pattern of size $L \times L$.

---
[a] *E-mail:* piaser@uni.opole.pl

When $S_\Delta(\varphi) = 0$ the distribution is perfectly homogeneous at a given scale $k$ while $S_\Delta(\varphi) = -k^2 [\varphi \ln \varphi + (1-\varphi) \ln(1-\varphi)]$ corresponds to its maximal spatial inhomogeneity. To apply that formula the knowledge of $F_{n_i}(\varphi)$ is needed. Such opportunity offers the coarsened lattice model of random two-phase systems [7] with a given grain size distribution (GSD). For grains of high $\{\sigma^H\}$ and low $\{\sigma^L\}$ conductivity, $\{4:2:1\}:\{1\}$ GSD was investigated, respectively. According this notation a grain of size '4' ('2') is considered as composed of four (two) grains of size '1'. All the grains occupy "positions" in model cells with the related probabilities, $p_4(\varphi)$, $p_2(\varphi)$, $p_1(\varphi)$ and $q_1(\varphi)$. By use of the model parameter $A \equiv p_2/p_1$, $\{1\}:\{1\}$ GSD evolves via $\{2:1\}:\{1\}$ towards $\{2\}:\{1\}$, see Fig. 1a.

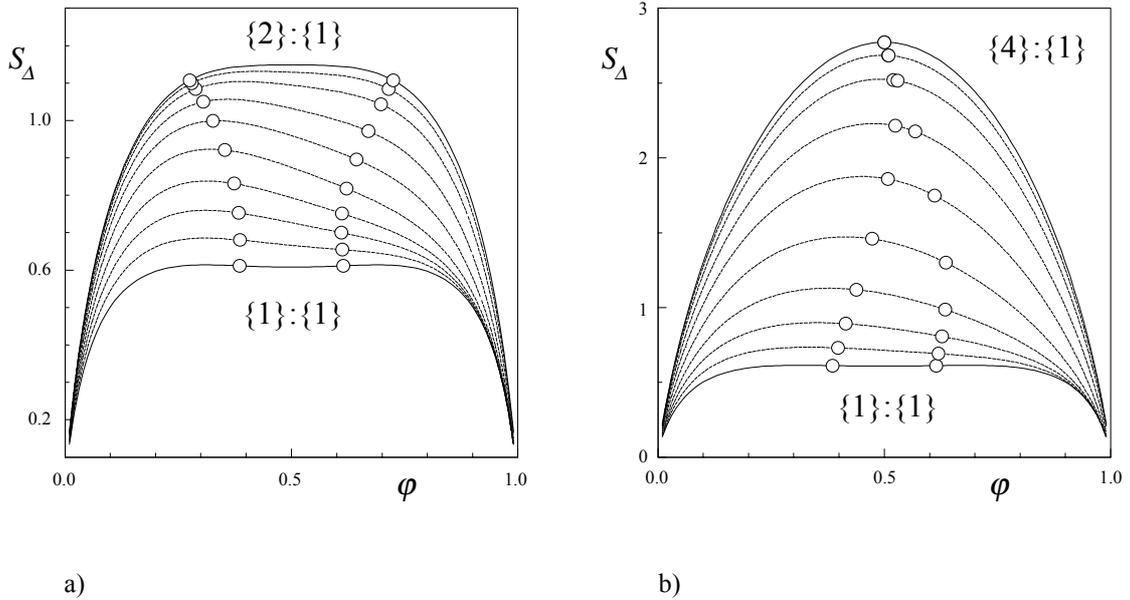

a)  b)

Fig. 1. Thermodynamic limit spatial disorder quantified by $S_\Delta(\varphi)$ for the coarsened lattice model [7] with $k = 2$. a) For $A = 0$ (bottom solid line), 0.1, 0.25, 0.5, 1, 2, 4, 10, 30 referring to $\{2:1\}:\{1\}$ GSDs (dashed lines) and $A \to \infty$ (top solid line). b) Same for $E$ with the dashed lines related to $\{4:1\}:\{1\}$ GSDs. In both cases the open left (right) circles correspond to the conductor-superconductor $p_c^I$ (conductor-insulator $p_c^{II}$) thresholds.

Similarly, for $E \equiv p_4/p_1$, $\{1\}:\{1\} \rightarrow \{4:1\}:\{1\} \rightarrow \{4\}:\{1\}$, see Fig. 1b (for more details the reader is referred to [7] where the non-monotonic behaviour of the conductor-superconductor $p_c^{I}$ and conductor-insulator $p_c^{II}$ threshold has been found).

Concluding, for $0 < A, E < \infty$ the broken topological equivalence of H- and L-phases causes the lack of invariance of the measure $S_\Delta(\varphi)$ under the replacement of $\varphi \leftrightarrow 1 - \varphi$ in contrast to the property of $S_\Delta(k)$ [7]. This is illustrated in Figs. 1a and 1b by the asymmetry of dashed curves regarding $\varphi = 0.5$, for which $p_c^{I} + p_c^{II} \neq 1$, i.e. the two thresholds are not symmetrical. In turn for limit GSDs the symmetric solid lines show at $\varphi = 0.5$ a shallow minimum for $\{1\}:\{1\}$, weak maximum for $\{2\}:\{1\}$ and distinct peak for $\{4\}:\{1\}$.